\documentclass[english,prd,twocolumn,nofootinbib,preprintnumbers]{revtex4}
\usepackage[latin1]{inputenc}
\usepackage{graphicx}
\usepackage{amsmath,amsthm,amssymb}
\usepackage{enumitem}
\usepackage{bm}

\usepackage{amsfonts}
\usepackage{dcolumn}
\usepackage{hyperref}

\def\be{\begin{equation}}
\def\ee{\end{equation}}
\def\ba{\begin{eqnarray}}
\def\ea{\end{eqnarray}}
\def\bs{\begin{subequations}}
\def\es{\end{subequations}}

\newcommand{\V}{{\text{\tiny $V$}}}

\usepackage{color}

\makeatother

\usepackage{babel}
\makeatother
\begin{document}

\title{Observational tests of inflation with a field derivative coupling to gravity}

\author{Shinji Tsujikawa}

\affiliation{Department of Physics, Faculty of Science, 
Tokyo University of Science, 
1-3, Kagurazaka, Shinjuku-ku, Tokyo 162-8601, Japan}

\begin{abstract}

A field kinetic coupling with the Einstein tensor leads to a gravitationally 
enhanced friction during inflation, by which even steep potentials with 
theoretically natural model parameters
can drive cosmic acceleration.  
In the presence of this non-minimal derivative coupling we place 
observational constraints on a number of representative inflationary 
models such as chaotic inflation, inflation with exponential potentials, 
natural inflation, and hybrid inflation.
We show that most of the models can be made compatible with 
the current observational data mainly due to 
the suppressed tensor-to-scalar ratio.

\end{abstract}

\date{\today}

\pacs{98.80.Cq, 95.30.Cq}

\maketitle

\section{Introduction}

Inflation has been the backbone of the high-energy cosmology 
over the past 3 decades \cite{infpapers}. 
The most simple source for inflation 
is a minimally coupled scalar field $\phi$ (``inflaton'') 
with a slowly varying potential $V(\phi)$ \cite{newinf,chaotic}
The spectra of density perturbations generated from 
the quantum fluctuations of inflaton are consistent with 
the temperature anisotropies observed in 
the Cosmic Microwave Background (CMB) \cite{infper}.

{}From the amplitude of the observed CMB 
anisotropies \cite{COBE,WMAP7}
the typical mass scale of inflation is known to be around 
$m \sim 10^{14}$~GeV \cite{Liddle}.
This is much larger than the electroweak scale 
($\sim 10^2$\,GeV), which suggests the requirement of 
new physics beyond the Standard Model of 
particle physics \cite{Riotto}.
In other words, for the potential 
$V(\phi)=(\lambda/4) (\phi^2-v^2)^2$ with 
$v \sim 10^2$~GeV, the coupling $\lambda$ 
is constrained to be $\lambda \sim 10^{-13}$ from 
the CMB normalization \cite{Liddle}, 
but this is much smaller than 
the coupling constant $\lambda \sim 0.1$
of the Higgs boson \cite{particlereview}.

There have been attempts to accommodate 
the Higgs field for inflation. 
One is to use a non-minimal field coupling 
$\xi R \phi^2/2$ with the Ricci scalar $R$ \cite{Bezrukov}
(see also Refs.~\cite{Maeda}).
If $\xi \gg 1$ the self coupling $\lambda$ can 
be as large as $\lambda \approx 10^{-10}\xi^2$
from the CMB normalization \cite{Salopek}.
Although this scenario is attractive, it is plagued by 
the unitary-violation problem associated with 
graviton exchange in $2 \to 2$ scalar scattering
around the energy scale $\Lambda_c \approx M_{\rm pl}/\xi$
(where $M_{\rm pl}=2.44 \times 10^{18}$~GeV is the 
reduced Planck mass) \cite{Burgess}.
Since $\Lambda_c$ is around the energy scale of inflation, 
some strong coupling effect
can give rise to additional corrections to the inflaton potential.

Another attempt is to employ a field derivative 
coupling with the Einstein tensor $G^{\mu \nu}$, i.e.
$G^{\mu \nu} \partial_{\mu} \phi \partial_{\nu} \phi/(2M^2)$,
where $M$ is a constant having a dimension of 
mass \cite{Germani1} (see also Ref.~\cite{Amendola} 
for the original work).
In the regime where the Hubble parameter $H$ 
is larger than $M$ the field evolves more slowly 
relative to the case of standard inflation 
due to a gravitationally enhanced friction.
Hence it is possible to reconcile steep  
potentials such as $V(\phi)=\lambda \phi^4/4$ ($\lambda \sim 0.1$)
with the CMB observations.

In Refs.~\cite{Germani1,Germaniper,Germani2,Watanabe,Germani11}
it was shown that, for a slow-rolling scalar field satisfying the 
condition $\varepsilon \equiv (\partial \phi)^2/(M^2 M_{\rm pl}^2) \ll 1$,
the strong coupling scale $\Lambda_c$ of the derivative coupling theory
is around $M_{\rm pl}$ in a homogeneous 
and isotropic cosmological background.
Provided that $H$ and $M$ are below the Planck scale, 
the theory is in a weak coupling regime 
with suppressed quantum corrections.

The property of the high cut-off scale $\Lambda_c$
around $M_{\rm pl}$ is associated with the fact that whenever 
the non-minimal derivative coupling to gravity dominates over the 
canonical kinetic term 
the theory possesses an asymptotic local shift symmetry for
$\varepsilon \ll 1$ \cite{Germani11}. 
This symmetry is related to the Galilean 
symmetry $\phi \to \phi+c+c_\mu x^{\mu}$ in Minkowski
space-time \cite{Nicolis}, 
but the difference is that the coordinate $x^{\mu}$ 
in the derivative coupling theory on curved 
backgrounds is linked to the covariantly constant 
Killing vectors \cite{Germani11}.
In the presence of a slowly varying inflaton potential 
such a local symmetry is only softly broken, so that the potential 
can be protected against quantum corrections during inflation.
The field self-interaction of the form 
$(\partial \phi)^2 \square \phi$ \cite{Nicolis,covaga}, which satisfies 
the Galilean symmetry in the limit of Minkowski space-time, 
also leads to the slow evolution of $\phi$ along the inflaton 
potential \cite{Kamada}
(see also Refs.~\cite{Galileoninf}).

A nice feature of the non-minimal derivative coupling 
with the Einstein tensor is that the mechanism of the gravitationally 
enhanced friction works for general steep potentials.
For instance, let us consider the potential of natural inflation, 
$V(\phi)=\Lambda^4 [1+\cos(\phi/f)]$, where $f$ characterizes 
the scale of the breaking of a global shift symmetry \cite{Freese}.
In order for this potential to be consistent with the CMB observations, 
we require that $f$ is larger than $3.5M_{\rm pl}$ in conventional 
slow-roll inflation \cite{Savage}.
Then the global symmetry is broken above the quantum 
gravity scale, in which case
quantum field theory may be invalid \cite{naturalpro}.
In the presence of the non-minimal derivative coupling to gravity, 
however, the scale $f$ can be much smaller than 
$M_{\rm pl}$ because of the gravitationally enhanced 
friction \cite{Germani2,Watanabe}.

Another example is the exponential potential 
$V(\phi)=V_0 e^{\beta \phi/M_{\rm pl}}$, 
whose dominance leads to the power-law expansion of the Universe 
(with the scale factor $a \propto t^{2/\beta^2}$, where $t$ is 
cosmic time) \cite{earlyexp}.
In higher-dimensional gravitational theories, exponential potentials 
often arise as the curvature of internal spaces related 
with the geometry of extra dimensions \cite{expmoti}.
In such cases the constant $\beta$ is usually larger than the order 
of unity, so that it is difficult to realize sufficient amount of inflation.
As we will see later, this problem can be circumvented 
by taking into account the non-minimal derivative coupling.
Moreover, unlike the standard case, inflation comes to end
with gravitational particle production.

In order to test the viability of inflationary models with the non-minimal
derivative coupling it is important to estimate the power spectra of density 
perturbations relevant to the CMB anisotropies.
In Refs.~\cite{Germaniper,Watanabe} the authors computed 
the inflationary observables such as the scalar spectral 
index $n_{\rm s}$, the tensor-to-scalar ratio $r$, and 
the nonlinear parameter $f_{\rm NL}^{\rm equil}$ of the 
equilateral scalar non-Gaussianities (see also 
Refs.~\cite{KYY,Gao,DT11}).
Since the scalar propagation speed is close to the 
speed of light during inflation, the scalar non-Gaussianities
are suppressed to be small ($|f_{\rm NL}^{\rm equil}| \ll 1$).
Hence $n_{\rm s}$ and $r$ are the two main observables to 
distinguish between different inflaton potentials.

In this paper we shall place observational constraints 
on a number of representative inflationary models in the 
presence of the field derivative coupling with the Einstein tensor.
We use the bounds derived from the joint data analysis 
of WMAP7 \cite{WMAP7}, Baryon Acoustic 
Oscillations (BAO) \cite{BAO}, 
and the Hubble constant measurement (HST) \cite{HST}.
Note that some constraints on Higgs inflation and natural inflation 
have been discussed in Refs.~\cite{Germaniper,Watanabe} without 
the CMB likelihood analysis.
In Ref.~\cite{Popa} the author carried out the 
cosmological Monte-Carlo simulation to
test Higgs inflation with the field derivative coupling to gravity.
Our analysis based on the recent observational data
is general enough to cover a wide variety 
of models such as chaotic inflation \cite{chaotic},  
inflation with exponential potentials \cite{earlyexp}, 
natural inflation \cite{Freese}, 
and hybrid inflation \cite{hybrid}.  
We show that the gravitationally enhanced 
friction mechanism can make most of the models 
compatible with the current observations.

\section{Background dynamics}
\label{secback}

We start with the following 4-dimensional action 
\be
S=\int d^{4}x\sqrt{-g}\left[ 
\frac{M_{\rm pl}^{2}}{2}R
-\frac12 \Delta^{\mu \nu} \partial_{\mu} \phi
\partial_{\nu} \phi -V(\phi) \right]\,,
\label{action}
\ee
where 
\be
\Delta^{\mu \nu}=g^{\mu \nu}
-\frac{1}{M^2}G^{\mu \nu}\,.
\label{Delta}
\ee
Here $g$ is a determinant of the 
space-time metric $g_{\mu \nu}$, 
$R$ is the Ricci scalar, 
$G^{\mu \nu}$ is the Einstein tensor, $M$ is a constant 
having a dimension of mass, and $V(\phi)$ is the potential 
of a scalar field $\phi$.

The action (\ref{action}) belongs to a class of the 
most general scalar-tensor theories having second-order
equations of motion (which is required to avoid the Ostrogradski 
instability) \cite{Horndeski,Deffayet,Char,KYY}. 
The Lagrangian in such general Horndeski's theories
is the sum of the terms ${\cal L}_2=K(\phi,X)$, 
${\cal L}_3=-G_3(\phi, X) \square \phi$, 
${\cal L}_4=G_4(\phi, X)R+G_{4,X} \times 
[{\rm field~derivative~terms}]$, and 
${\cal L}_5=G_5(\phi, X)G^{\mu \nu} (\nabla_{\mu} \nabla_{\nu} \phi)
-(G_{5,X}/6) \times [{\rm field~derivative~terms}]$, where 
$K$, $G_i$ ($i=3,4,5$) are functions of 
$\phi$ and $X=-g^{\mu \nu} \partial_{\mu} \phi \partial_{\nu} \phi/2$, 
and $G_{i,X}=\partial G_i/\partial X$
\cite{Deffayet,KYY}.
The conditions for the avoidance of ghosts and Laplacian instabilities
were derived in Refs.~\cite{KYY,DTjcap}.
These conditions can be used to restrict the
functional forms of $K$, $G_i$ ($i=3,4,5$) to construct theoretically 
consistent models of inflation.

The non-minimal derivative coupling in Eq.~(\ref{action}) is recovered 
in the Horndeski's Lagrangian by choosing the function $G_{5}=-\phi/(2M^2)$
after integration by parts.
The sign in front of the term $G^{\mu \nu}/M^2$
in Eq.~(\ref{Delta}) is chosen to 
avoid the appearance of ghosts in the scalar 
sector \cite{Germani1,Germani2}. 

In Ref.~\cite{Germani11} it was found that in a manifold 
having integrable (covariantly constant) Killing vectors $\xi^{a}$
the field Lagrangian 
$-\Delta^{\mu \nu} \partial_{\mu} \phi \partial_{\nu} \phi/2$ 
in Eq.~(\ref{action}) is invariant under the (curved-space) 
Galilean transformation $\phi(x) \to \phi(x)+c+c_a \int^{x}_{x_0} \xi^{a}$, 
where $c$, $c_a$, $x_0$ are constants and $x$ is a space-time
coordinate. The existence of the Galilean symmetry has an advantage
that the theory can be quantum mechanically under control \cite{Trod}. 

Imposing the above Galilean symmetry in the curved background
with integrable Killing vectors, Germani {\it et al.} \cite{Germani11}
showed that the second-order Lagrangians are restricted to take
the forms $-\Delta^{\mu \nu} \partial_{\mu} \phi \partial_{\nu} \phi/2$
or $X \square \phi$ (plus a field derivative coupling with 
the double dual Riemann tensor).
In the small derivative regime in which the condition 
$(\partial \phi)^2/(M^2 M_{\rm pl}^2) \ll 1$ is satisfied 
(e.g., during inflation), an approximate infinitesimal shift symmetry
$\phi \to \phi+f(x)$ (where $f(x)$ is an arbitrary function of space-time
coordinates $x$) emerges for the Lagrangian 
$G^{\mu \nu} \partial_{\mu} \phi \partial_{\nu} \phi/(2M^2)$, 
provided that the metric is shifted appropriately \cite{Germani11}.
The existence of such a gauge symmetry can allow 
the theory (\ref{action}) to be protected against quantum 
corrections even up to the Planck scale.
Note that the term $X \square \phi$ does not possess
such a general gauge shift symmetry.

The scale of unitarity violation for the theory (\ref{action}) was 
estimated in Ref.~\cite{Germani1} in the context of Higgs inflation.
In Standard Model we can consider the $\phi \phi \to \phi \phi$ 
scattering via graviton exchange, where $\phi$ is one of the real 
scalar degrees of freedom for the Higgs doublet.
We expand the metric in the form $g_{\mu \nu}=
g_{\mu \nu}^{(0)}+h_{\mu \nu}/M_{\rm pl}$, where 
$g_{\mu \nu}^{(0)}=(-1, a^2(t), a^2(t), a^2(t))$ is the metric on
the flat Friedmann-Lema\^{i}tre-Robertson-Walker
(FLRW) background 
($a(t)$ is the scale factor with cosmic time $t$).
We are interested in the high-friction regime in which 
the Hubble parameter $H=\dot{a}/a$ 
(a dot represents a derivative with respect to $t$)
is much larger than $M$.
In this regime the field $\phi$ is expanded as 
$\phi=\phi_0+M\chi/(\sqrt{3}H)$, where $\chi$ is 
a canonically normalized field perturbation.
The first non-renormalizable operator associated with 
the interaction between gravitons and scalars 
is given by \cite{Germani1}
\be
I=\frac{1}{2H^2 M_{\rm pl}} \partial^2 h^{\mu \nu}
\partial_{\mu} \chi \partial_{\nu} \chi\,.
\label{inter}
\ee
A power counting analysis gives the unitary bound 
$\Lambda \simeq (2H^2 M_{\rm pl})^{1/3}$.
For the suppression of higher dimensional operators
we require the condition $R<\Lambda^2$.
On using the relation $R \simeq 12H^2$
this condition translates into $H<5 \times 10^{-2} M_{\rm pl}$, 
which is satisfied during inflation.

The discussion of the unitary bound given above can 
be applied to the multi-field inflationary models in which one of the 
fields is not necessarily responsible for the cosmic acceleration.
In single field models where only one field $\phi$
leads to inflation the unitary bound can be
as close as the Planck scale $M_{\rm pl}$
in the regime where the condition 
$(\partial \phi)^2/(M^2 M_{\rm pl}^2) \ll 1$ 
is satisfied \cite{Watanabe,Germani11}.
In this case the slow-roll evolution of the field $\phi$
suppresses the interaction (\ref{inter}) below
the Planck scale.

In the following let us study the background dynamics
for the theory described by the action (\ref{action}). 
In the flat FLRW background the
equations of motion following from 
the action (\ref{action}) are 
\ba
& & 3 M_{\rm pl}^2 H^2=
\frac12 \dot{\phi^2} \left(1+9 \frac{H^2}{M^2}
\right)+V(\phi)\,,
\label{Hubbleeq} \\
& & \frac{1}{a^3} \frac{d}{dt} \left[ 
a^3 \dot{\phi} \left( 1+3 \frac{H^2}{M^2} 
\right) \right]+V_{,\phi}=0\,,
\label{Jeq}
\ea
where $V_{,\phi}=dV/d\phi$.
In order to solve the dynamical equations numerically,
it is convenient to introduce the following 
dimensionless variables
\be
x=\frac{\phi}{M_{\rm pl}}\,,\qquad
y=\frac{\dot{\phi}}{MM_{\rm pl}}\,,\qquad
z=\frac{H}{M}\,.
\ee
Differentiating Eq.~(\ref{Hubbleeq}) with respect to 
$t$ and using Eq.~(\ref{Jeq}) to eliminate $\dot{H}$, 
we obtain the second-order equation for the field $\phi$.
It then follows that 
\ba
\frac{dx}{d\tau} &=& y\,,
\label{dxeq} \\
\frac{dy}{d \tau} &=& 
-\frac{3yz (2-3y^2)(1+3z^2)+(2-y^2)\hat{V}_{,\phi}(x)}
{2(1+3z^2)+y^2 (9z^2-1)}\,,
\label{dyeq} \\
z &=& \sqrt{\frac{y^2+2\hat{V}(x)}{6-9y^2}}\,,
\label{zeq}
\ea
where
\be
\tau=M t\,,\qquad
\hat{V}_{,\phi}=\frac{V_{,\phi}}{M^2 M_{\rm pl}}\,,
\qquad
\hat{V}=\frac{V}{M^2 M_{\rm pl}^2}\,.
\ee

We are interested in slow-roll inflation in which 
cosmic acceleration is mainly driven by 
the potential energy $V(\phi)$. 
In this case Eqs.~(\ref{Hubbleeq}) and 
(\ref{Jeq}) reduce to 
\ba
& & 3M_{\rm pl}^2 H^2 \simeq V(\phi)\,,
\label{sloweq1} \\
& & 3H {\cal A}\,\dot{\phi} +V_{,\phi}
\simeq 0\,,
\label{sloweq2}
\ea
where 
\be
{\cal A}=1+3\frac{H^2}{M^2}\,.
\label{calA}
\ee
We define the following slow-roll parameters
\ba
& & \epsilon=-\frac{\dot{H}}{H^2}\,,\qquad
\delta_{\phi}=\frac{\ddot{\phi}}{H \dot{\phi}}\,,
\nonumber \\
& &
\delta_{X}=\frac{\dot{\phi}^2}{2H^2 M_{\rm pl}^2}\,,
\qquad
\delta_D=\frac{\dot{\phi}^2}{4M^2 M_{\rm pl}^2}\,.
\ea
For the validity of the slow-roll approximation 
we require that $\{ \epsilon, |\delta_{\phi}|, 
\delta_{X}, \delta_{D} \} \ll 1$.
Taking the time-derivative of Eq.~(\ref{sloweq1})
and using Eq.~(\ref{sloweq2}), we have 
\be
\epsilon \simeq \delta_{X}+6\delta_{D}
\simeq \frac{\epsilon_{\V}}{{\cal A}}\,,
\label{epre}
\ee
where
\be
\epsilon_{\V} = \frac{M_{\rm pl}^2}{2}
\left( \frac{V_{,\phi}}{V} \right)^2\,.
\label{epV}
\ee
This shows that $\epsilon \ll \epsilon_{\V}$
for ${\cal A} \gg 1$ and hence
the evolution of the field $\phi$ slows down 
relative to that in standard slow-roll inflation.

The field value $\phi_f$ at the end of inflation is 
known by solving $\epsilon (\phi_f)=1$, i.e.
\be
\epsilon_{\V} (\phi_f) \left[ 1+\frac{V(\phi_f)}
{M^2 M_{\rm pl}^2} \right]^{-1}=1\,.
\label{epV2}
\ee
The number of e-foldings from the time $t$ during inflation to 
the time $t_f$ at the end of inflation is defined by 
$N=\int_{t}^{t_f}H(\tilde{t})\,d\tilde{t}$.
On using Eqs.~(\ref{sloweq1}) and (\ref{sloweq2}),  
it follows that 
\be
N \simeq \frac{1}{M_{\rm pl}^2} \int_{\phi_f}^{\phi}
\left( 1+\frac{V}{M^2 M_{\rm pl}^2} \right)
\frac{V}{V_{,\tilde{\phi}}} d\tilde{\phi}\,.
\label{efold}
\ee
If ${\cal A} \gg 1$ (i.e. $H^2 \gg M^2$) during inflation, 
we can neglect the first term inside the bracket of 
Eq.~(\ref{efold}) relative to the second one.
This is not the case for inflation in which
the transition from the regime $H>M$ to 
the regime $H<M$ occurs prior to the onset of 
reheating.

\begin{figure}
\includegraphics[height=3.2in,width=3.4in]{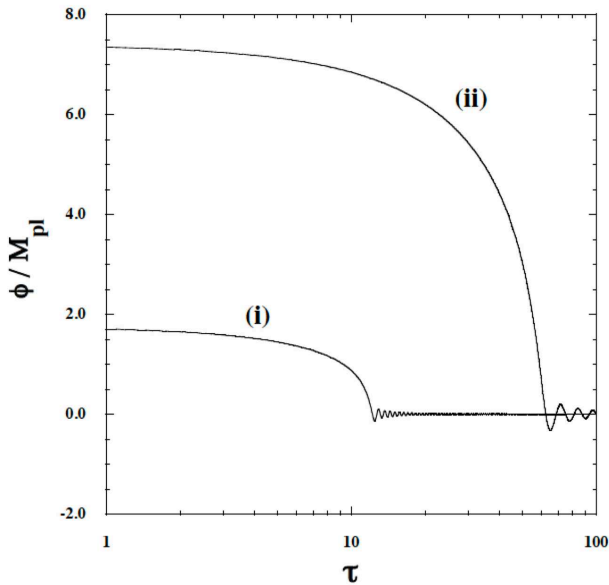}
\caption{Evolution of the field $\phi$ versus $\tau=Mt$
for (i) $\alpha=100$ and (ii) $\alpha=0.25$.
The initial conditions at $N=60$ are chosen 
by using the slow-roll equations (\ref{sloweq1}) and 
(\ref{sloweq2}), i.e., 
(i) $x=1.757$, $y=-0.526$, and
(ii) $x=7.397$, $y=-0.521$.
\label{fig1}}
\end{figure}

As an example, let us consider chaotic 
inflation \cite{chaotic} with the potential 
\be
V(\phi)=\frac{\lambda}{n} \phi^n\,,
\ee
where $\lambda$ and $n$ are constants.
In this case Eqs.~(\ref{efold}) and (\ref{epV2}) read
\ba
\hspace{-0.9cm}
& & N = \frac{x^2}{2n} \left[ 1+\frac{2\alpha}
{n(n+2)} x^n \right]-\frac{x_f^2}{2n} \left[ 1+\frac{2\alpha}
{n(n+2)} x_f^n \right],
\label{chaotinre} \\
\hspace{-0.9cm}
&& 2 x_f^2 \left( 1+\frac{\alpha}{n} x_f^n \right)
=n^2\,,
\label{chaotinre2}
\ea
where 
\be
\alpha=\frac{\lambda M_{\rm pl}^{n-2}}{M^2}\,,
\qquad
x_f=\frac{\phi_f}{M_{\rm pl}}\,.
\ee
For the quadratic potential $V(\phi)=m^2 \phi^2/2$
(i.e. $\lambda=m^2$ and $n=2$) 
one has $x_f^2=(\sqrt{1+4\alpha}-1)/\alpha$ and 
\be
x^2=\left( \sqrt{2+2\sqrt{1+4\alpha}+4\alpha(4N+1)}
-2 \right)/\alpha\,,
\label{xslow}
\ee
where $\alpha=m^2/M^2$.
In the General Relativistic (GR) limit ($\alpha \to 0$) this gives $x^2 \to 4N+2$.
In the high-friction limit ($\alpha \to \infty$) one has
$x^2 \to 2\sqrt{(4N+1)/\alpha}$, 
which means that the field value is 
smaller than that in standard chaotic inflation.

In order to confirm the accuracy of the slow-roll approximation 
we solve the full equations of motion 
(\ref{dxeq})-(\ref{zeq}) numerically.
In Fig.~\ref{fig1} the evolution of the field $\phi$
is plotted for the potential $V(\phi)=m^2 \phi^2/2$, i.e.
$\hat{V}_{,\phi}=\alpha x$ and $\hat{V}=\alpha x^2/2$
in Eqs.~(\ref{dyeq}) and (\ref{zeq}).
We choose the initial conditions of $x$ and $y$ at $N=60$ 
by using the values derived under the slow-roll approximation.

The numerical simulations labeled as (i) and (ii) 
in Fig.~\ref{fig1} correspond to the parameters
$\alpha=100$ and $\alpha=0.25$, respectively.
In the case (i) the numerical value of $z=H/M$ at the end 
of inflation ($x_f=0.44$) is $z_f=2.2$, which means that 
the solution is in the high-friction regime during inflation.
In this case $H$ drops below $M$
at the reheating stage.
After inflation there is a transient period with $H>M$
in which the slow-roll condition is violated.
In this regime some quantum corrections may come into 
play to the action (\ref{action}).
As long as such corrections are unimportant in
the field equations (\ref{Hubbleeq}) and (\ref{Jeq}),  
we find that the inflaton oscillation 
is not disturbed during the transient period
(see Fig.~\ref{fig1}).
In the case (ii) we have $z_f=0.3$ at 
$x_f=1.3$ and hence the system enters 
the regime $H<M$ during inflation.
In this case the qualitative behavior for the oscillation 
of inflaton at reheating is not much 
different from that in standard inflation.

We confirmed that the difference between
the numerical and analytic values of the number of 
e-foldings acquired during inflation 
is usually less than a few percent.
This shows that the slow-roll approximation 
employed in Eqs.~(\ref{sloweq1}) and (\ref{sloweq2})
can be trustable.

\section{The spectra of density perturbations}
\label{formulas}

The spectra of scalar and tensor perturbations generated 
in the theories given by the action (\ref{action}) were derived
in Refs.~\cite{Germaniper,KYY,Watanabe,DT11}.
Here we briefly review their formulas in order to apply them 
to concrete inflationary models in Sec.~\ref{obsercon}.

The perturbed metric about the flat FLRW 
background is given by \cite{permet}
\ba
ds^2 &=& -(1+2A)dt^2+2\partial_i B\,dt\,dx^i 
\nonumber \\
& & +a^2(t)\,\left[(1+2{\cal R})\delta_{ij}+h_{ij}
\right] dx^i dx^j\,,
\label{permet}
\ea
where $A$, $B$, and ${\cal R}$ are scalar metric perturbations, 
and $h_{ij}$ are tensor perturbations which are 
transverse and traceless.
The spatial part of a gauge-transformation vector $\xi^{\mu}$ 
is fixed by gauging away a perturbation $E$ appearing
as a form $E_{,ij}$ in the metric (\ref{permet}).
We decompose the inflaton field into the background and 
inhomogeneous parts, as $\phi=\phi_{0}(t)
+\delta \phi (t, {\bm x})$.
In the following we choose the uniform-field gauge 
characterized by $\delta \phi=0$, which fixes the 
time-component of the vector $\xi^{\mu}$.

Expanding the action (\ref{action}) up to second order
in perturbations and using the Hamiltonian and momentum 
constraints, we obtain the second-order action
for scalar perturbations \cite{KYY,DT11}
\be
S_{\rm s}^{(2)}=\int dt\,d^{3}x\, a^{3}
Q_{\rm s} \left[\dot{{\cal R}}^{2}
-\frac{c_{\rm s}^{2}}{a^{2}}\,(\partial{\cal R})^{2}\right]\,,
\ee
where 
\ba
\hspace{-0.3cm}
Q_{\rm s} & = & \frac{w_{1}(4w_{1}w_{3}+9w_{2}^{2})}{3w_{2}^{2}}\,,
\label{Qdef} \\
\hspace{-0.3cm}
c_{\rm s}^{2} & = & \frac{3(2w_{1}^{2}w_{2}H-w_{2}^{2}w_{4}
+4w_{1}\dot{w}_{1}w_{2}-2w_{1}^{2}\dot{w}_{2})}
{w_{1}(4w_{1}w_{3}+9w_{2}^{2})},
\ea
and 
\ba
& & w_1=M_{\rm pl}^2 (1-2\delta_D)\,,\quad
w_2=2 H M_{\rm pl}^2 (1-6\delta_D)\,,\nonumber \\
& & w_3=-3H^2 M_{\rm pl}^2 (3-\delta_{X}
-36 \delta_{D})\,,\nonumber \\
& & w_4=M_{\rm pl}^2 (1+2\delta_{D})\,.
\ea
In order to avoid the appearance of scalar ghosts and 
Laplacian instabilities we require that 
$Q_{\rm s}>0$ and $c_{\rm s}^2>0$.
Picking up the dominant contributions to $Q_{\rm s}$ and 
$c_{\rm s}^2$ under the slow-roll approximation, 
we obtain
\ba
Q_{\rm s} &\simeq& M_{\rm pl}^2 
(\delta_{X}+6\delta_{D}) \simeq 
M_{\rm pl}^2\,\epsilon \simeq M_{\rm pl}^2
\frac{\epsilon_{\V}}{{\cal A}}\,,\label{Qs} \\
c_{\rm s}^2 &\simeq& 1-
\frac{2\delta_D (3\delta_X+34\delta_D-2\delta_{\phi})}
{\delta_{X}+6\delta_D}\,,
\label{cs}
\ea
which mean that $c_{\rm s}^2=1-{\cal O}(\epsilon)$.
The power spectrum of the curvature perturbation ${\cal R}$, 
which is evaluated at $c_{\rm s} k=aH$ ($k$ is a comoving 
wavenumber), is given by 
\be
{\cal P}_{\rm s}=\frac{H^2}{8\pi^2 Q_{\rm s}c_{\rm s}^3}
\simeq \frac{V^3}{12\pi ^2 M_{\rm pl}^6 V_{,\phi}^2} 
\left( 1+\frac{V}{M^2 M_{\rm pl}^2} \right)\,,
\label{COBE}
\ee
where in the last approximate equality we used
Eqs.~(\ref{sloweq1}), (\ref{calA}), (\ref{epV}), 
(\ref{Qs}), and (\ref{cs}).
The scalar spectral index is 
\ba
n_{\rm s}-1 &=& \frac{d \ln {\cal P}_{\rm s}}
{d \ln k}\biggr|_{c_{\rm s}k=aH} \nonumber \\
&\simeq& -\frac{1}{{\cal A}} 
\left[ 2\epsilon_{\V} \left( 4-\frac{1}{{\cal A}} \right)
-2\eta_{\V} \right]\,,
\label{ns}
\ea
where $\epsilon_{\V}$ is defined in Eq.~(\ref{epV}), 
and
\be
\eta_{\V}= M_{\rm pl}^2 \frac{V_{,\phi \phi}}{V}\,.
\ee
In the high-friction limit (${\cal A} \gg 1$) one has 
$n_{\rm s}-1 \simeq -(8\epsilon_{\V}-2\eta_{\V})/{\cal A}$
with ${\cal A} \simeq V/(M^2 M_{\rm pl}^2)$, whereas
$n_{\rm s}-1 \simeq -6\epsilon_{\V}+2\eta_{\V}$
in the GR limit (${\cal A} \simeq 1$).

The intrinsic tensor perturbation $h_{ij}$ can be 
decomposed into two independent polarization
modes, i.e. 
$h_{ij}=h_{+}e_{ij}^{+}+h_{\times}e_{ij}^{\times}$.
In Fourier space we normalize the two modes, as
$e_{ij}^{p}(\bm{k})e_{ij}^{p}(-\bm{k})^{*}=2$
(where $p={+},{\times}$) and 
$e_{ij}^{+}(\bm{k})e_{ij}^{\times}(-\bm{k})^{*}=0$.
Then the second-order action for 
tensor perturbations can be written as  \cite{KYY,DT11}
\be
S_{\rm t}^{(2)}=\sum_{p}\int dt\,d^{3}x\, 
a^{3}Q_{\rm t} \left[\dot{h}_{p}^{2}
-\frac{c_{\rm t}^{2}}{a^{2}}\,(\partial 
h_{p})^{2}\right]\,,
\ee
where 
\ba
Q_{\rm t} &=& w_1/4 =M_{\rm pl}^2 
(1-2\delta_{D})/4\,,\\
c_{\rm t}^2 &=& w_4/w_1 
=1+4\delta_{D}+{\cal O} (\epsilon^2)\,.
\ea
This shows that, unlike the scalar propagation speed squared,  
$c_{\rm t}^2$ is slightly larger than 1 during inflation. 
Since Lorentz invariance is explicitly 
broken on the FLRW background, the superluminal  
mode does not necessarily imply a violation of causality. 
The tensor power spectrum is given by 
\be
{\cal P}_{\rm t}=\frac{H^2}{2\pi^2 Q_{\rm t} c_{\rm t}^3}
\simeq \frac{2V}{3\pi^2 M_{\rm pl}^4}\,,
\ee
which is evaluated at $c_{\rm t}k=aH$.
The tensor spectral index is 
\be
n_{\rm t} = \frac{d \ln {\cal P}_{\rm t}}
{d \ln k}\biggr|_{c_{\rm t}k=aH} \simeq
-2\epsilon
\ee
The tensor-to-scalar ratio is 
\be
r = \frac{{\cal P}_{\rm t}}
{{\cal P}_{\rm s}}\biggr|_{k \simeq aH}  
\simeq \frac{16\epsilon_{\V}}
{{\cal A}} \simeq 16 \epsilon\,,
\label{ratio}
\ee
from which we obtain the consistency relation 
\be
r \simeq -8 n_{\rm t}\,.
\label{consistency}
\ee
This relation is the same as that in conventional 
inflation at leading order in slow-roll.

The running spectral indices 
$\alpha_{\rm s}=d n_{\rm s}/d \ln k|_{c_{\rm s}k=aH}$
and  $\alpha_{\rm t}=d n_{\rm t}/d \ln k|_{c_{\rm t}k=aH}$ are 
second order in slow-roll parameters.
They are set to be 0 in the CMB likelihood 
analysis. The consistency relation (\ref{consistency}) reduces the inflationary 
observables to three, i.e., $n_{\rm s}$, $r$, and ${\cal P}_{\rm s}$.
These observables are varied in the likelihood analysis
with the pivot wavenumber $k_0=0.002$\,Mpc$^{-1}$,
by assuming the flat $\Lambda$-cold-dark-matter model.

Since $c_{\rm s}^2=1-{\cal O}(\epsilon)$, the general formula
for the equilateral non-Gaussianities of scalar perturbations \cite{DT11}
shows that the nonlinear parameter $f_{\rm NL}^{\rm equil}$
is of the order of $\epsilon$ \cite{Watanabe}.
Hence the scalar non-Gaussianities do not provide additional 
constraints to those derived from the linear perturbations.

\section{Observational constraints}
\label{obsercon}

In the presence of the field derivative coupling to the Einstein 
tensor we place observational constraints on a number of models
such as (i) chaotic inflation, (ii) inflation with exponential 
potentials, (iii) natural inflation, and (iv) hybrid inflation.
Our analysis covers most of the representative 
inflaton potentials proposed in literature.

\subsection{Chaotic inflation}
\label{chaoticsec}

We start with chaotic inflation characterized 
by the potential 
\be
V(\phi)=\frac{\lambda}{n} \phi^{n}\,,
\ee
where, for $n=2$, we use the notation $\lambda=m^2$
as in the previous section.
Under the slow-roll approximation the dimensionless field 
$x=\phi/M_{\rm pl}$ is related with the number of 
e-foldings $N$ as Eq.~(\ref{chaotinre}).
{}From Eqs.~(\ref{ns}) and (\ref{ratio}) the scalar spectral 
index and the tensor-to-scalar ratio are given, 
respectively, by
\ba
n_{\rm s} &=& 1-\frac{n^2[n(n+2)+2(n+1)\alpha x^n]}
{x^2 (n+\alpha x^n)^2}\,,
\label{nschao} \\
r &=& \frac{8n^3}{x^2 (n+\alpha x^n)}\,,
\label{rchao}
\ea
where $\alpha=\lambda M_{\rm pl}^{n-2}/M^2$.

\begin{figure}
\includegraphics[height=3.3in,width=3.4in]{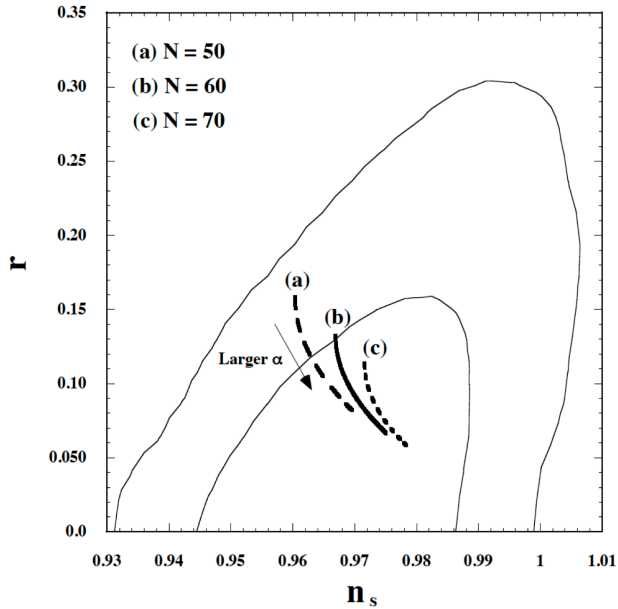}
\caption{Observational constraints on chaotic inflation 
with the quadratic potential $V(\phi)=m^2 \phi^2/2$
in the ($n_{\rm s}, r$) plane with the three different 
numbers of e-foldings ($N=50, 60, 70$).
We evaluate the theoretical values of $n_{\rm s}$ and $r$
in the range $10^{-8} \le \alpha=m^2/M^2 \le 10^{8}$.
The thin solid curves correspond to the 1$\sigma$ (inside)
and 2$\sigma$ (outside) observational contours 
constrained by the joint data analysis of WMAP7, BAO, and HST.
For larger values of $\alpha=m^2/M^2$ the tensor-to-scalar 
ratio $r$ gets smaller, whereas the scalar 
spectral index increases.
\label{fig2}}
\end{figure}

In the limit that $\alpha \to 0$ one has $x=\sqrt{2n (4N+n)}/2$
from Eqs.~(\ref{chaotinre}) and (\ref{chaotinre2}), 
in which case $n_{\rm s}$ and $r$ are
\ba
n_{\rm s} &=& 1-\frac{2(n+2)}{4N+n}\,,
\label{chao1d}
\\
r &=& \frac{16n}{4N+n}\,.
\label{chao1}
\ea
These values correspond to those for standard chaotic inflation.
If $N=60$, then $n_{\rm s}=0.967$, $r=0.132$ for $n=2$
and $n_{\rm s}=0.951$, $r=0.262$ for $n=4$.
In another limit $\alpha \to \infty$ one has 
$x^{n+2}=[2N(n+2)+n]n^2/(2\alpha)$ and 
\ba
n_{\rm s} &=& 1-\frac{4(n+1)}{2(n+2)N+n}\,,
\label{chao2d}\\
r &=& \frac{16n}{2(n+2)N+n}\,.
\label{chao2}
\ea
If $N=60$, then $n_{\rm s}=0.975$, $r=0.066$ for $n=2$
and $n_{\rm s}=0.972$, $r=0.088$ for $n=4$.

In the intermediate values of $\alpha$ between $(0, \infty)$
we need to solve Eq.~(\ref{chaotinre}) for $x$ by using 
Eq.~(\ref{chaotinre2}).
When $n=2$ the field value $x$ can be expressed as Eq.~(\ref{xslow}), 
in which case $n_{\rm s}$ and $r$ are known from Eqs.~(\ref{nschao}) 
and (\ref{rchao}) for given values of $\alpha=m^2/M^2$ and $N$.
In Fig.~\ref{fig2} we plot the theoretical values of $n_{\rm s}$
and $r$ for $n=2$ as a function of $\alpha$ 
(between $10^{-8} \le \alpha \le 10^{8}$) with three different 
values of $N~(=50, 60, 70)$, together with 
the $1\sigma$ and $2\sigma$ observational contours 
constrained by the joint 
data analysis of WMAP7 \cite{WMAP7}, BAO \cite{BAO}, 
and HST \cite{HST}.
For $\alpha=10^{-8}$ these observables are close to 
the values estimated by Eqs.~(\ref{chao1d}) and (\ref{chao1}), 
whereas for $\alpha=10^{8}$ they are close to those 
given by Eqs.~(\ref{chao2d}) and (\ref{chao2}).
For larger $\alpha$, $r$ gets smaller whereas 
$n_{\rm s}$ increases, so that the quadratic potential
shows better compatibility with observations
in the presence of the field derivative coupling to gravity.

\begin{figure}
\includegraphics[height=3.3in,width=3.4in]{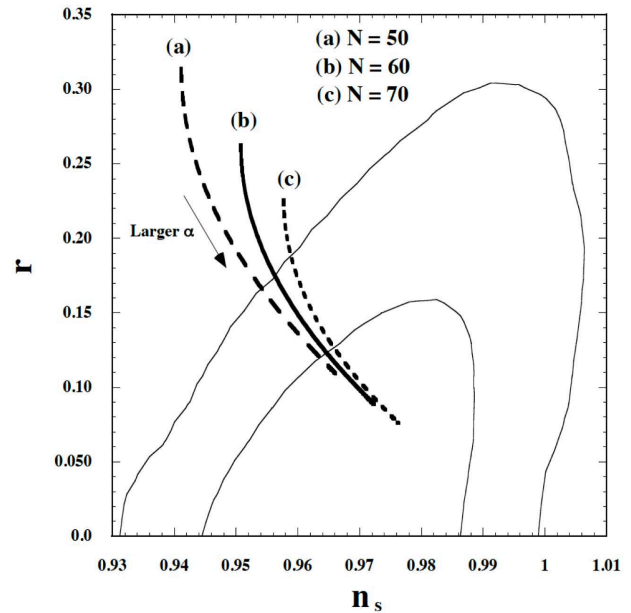}
\caption{Observational constraints on chaotic inflation 
with the quartic potential $V(\phi)=\lambda \phi^4/4$
in the range $10^{-8} \le \alpha=\lambda (M_{\rm pl}/M)^2 \le 10^{8}$ 
with the three different values of $N$.
The $1\sigma$ and $2\sigma$ observational contours 
are the same as those in Fig.~\ref{fig2}.
While the standard case ($\alpha=0$) is outside the 
$2\sigma$ bound, the field derivative coupling to 
gravity can make the quartic potential 
compatible with observations.
\label{fig3}}
\end{figure}

In Fig.~\ref{fig3} the theoretical values of $n_{\rm s}$
and $r$ are plotted for $n=4$ as a function of 
$\alpha=\lambda (M_{\rm pl}/M)^2$
between $(10^{-8}, 10^{8})$
with $N=50, 60, 70$.
In the limit that $\alpha \to 0$ the quartic potential
is outside the $2\sigma$ observational contour 
for $N$ smaller than 70. 
In the presence of the field derivative coupling to gravity
the model can be compatible with the current observations 
due to the suppressed tensor-to-scalar ratio and
the larger spectral index.
For $N=60$ the parameter $\alpha$ is constrained to be
\be
\alpha>3 \times 10^{-5}~~(95\,\%~{\rm CL})\,,
\ee
and $\alpha>4 \times 10^{-4}$ (68 \% CL).
For $N=50$ the constraints are
$\alpha>2 \times 10^{-4}$ (95 \% CL) and 
$\alpha>1 \times 10^{-2}$ (68 \% CL).

Using the scalar power spectrum (\ref{COBE}), the CMB 
normalization by WMAP \cite{WMAP7} corresponds to
\be
\frac{\alpha}{12 \pi^2 n^4} \left( \frac{M}{M_{\rm pl}} \right)^2
x_{60}^{n+2} \left( n+\alpha x_{60}^n \right) 
\simeq 2.4 \times 10^{-9}\,,
\label{CMBnor}
\ee
where $x_{60}$ is the value of $x$ at $N=60$.
In the regime $\alpha \gg 1$ one has 
$x^{n+2}=[2N(n+2)+n]n^2/(2\alpha)$, so that 
Eq.~(\ref{CMBnor}) gives
\ba
\frac{m}{M_{\rm pl}} &\simeq& 1.5 \times 10^{-10}\,
\frac{M_{\rm pl}}{M} \qquad {\rm for}~~n=2\,,\\
\lambda &\simeq& 5.9 \times 10^{-32}
\left( \frac{M_{\rm pl}}{M} \right)^4
\qquad {\rm for}~~n=4\,.
\ea
In the case of the quartic potential it is possible to 
realize $\lambda \simeq 0.1$ for 
$M \simeq 2.8 \times 10^{-8}\,M_{\rm pl}$.

\subsection{Exponential potentials}

Let us proceed to the exponential potential
\be
V(\phi)=V_0\,e^{\beta \phi/M_{\rm pl}}\,,
\ee
where $V_0$ and $\beta$ are constants.
In this case one has $\epsilon_{\V}=\beta^2/2$ and 
$\eta_{\V}=\beta^2$.
Hence in standard slow-roll inflation we require
the condition $\beta^2 \ll 1$.
Moreover this corresponds to the power-law inflation 
without the graceful exit \cite{earlyexp}.
In the presence of the field derivative coupling to gravity, 
however, the slow-roll parameter $\epsilon=\epsilon_{\V}/{\cal A}$ 
can be smaller than 1 even for steep exponential 
potentials with $\beta^2 \gtrsim 1$.
Inflation ends when $\epsilon$ grows to the order 
of unity, which is followed by reheating 
with gravitational particle production. 
This situation is analogous to that in braneworld inflation 
where the dominance of the density squared term
($\rho^2$) can lead to cosmic acceleration for steep
exponential potentials \cite{Copeland}.

\begin{figure}
\includegraphics[height=3.3in,width=3.4in]{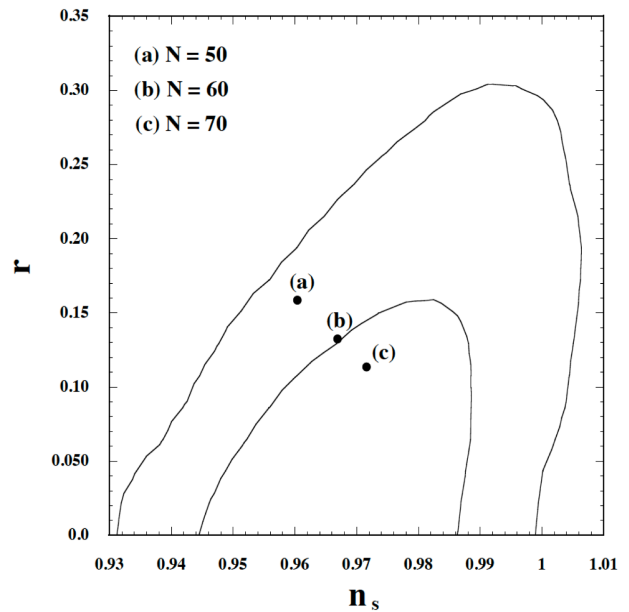}
\caption{Observational constraints on inflation with 
the exponential potential $V(\phi)=V_0 e^{\beta \phi/M_{\rm pl}}$ 
in the regime ${\cal A} \gg 1$
for $N=50, 60, 70$. The $1\sigma$ and $2\sigma$ 
observational contours are the same as 
those in Fig.~\ref{fig2}.
Even the steep exponential potentials with $\beta$ 
larger than 1 can be allowed from the current 
observational data. 
\label{fig4}}
\end{figure}

We focus on the case in which the condition 
$V/(M^2 M_{\rm pl}^2) \gg 1$ is satisfied during 
the whole stage of inflation.
Using Eq.~(\ref{epV2}) the field value $\phi_f$
at the end of inflation can be estimated as
\be
\frac{V_0 e^{\beta 
\phi_f/M_{\rm pl}}}{M^2 M_{\rm pl}^2}
\simeq \frac{\beta^2}{2}\,,
\ee
which implies that $\beta^2 \gg 1$.
{}From Eq.~(\ref{efold}) it follows that 
\be
\frac{V_0 e^{\beta \phi/M_{\rm pl}}}
{\beta^2 M^2 M_{\rm pl}^2} \simeq 
N+\frac12\,.
\ee
In the regime ${\cal A} \gg 1$ the scalar spectral index 
(\ref{ns}) and the tensor-to-scalar
ratio (\ref{ratio}) reduce to  
\ba
n_{\rm s} &\simeq& 1-\frac{4}{2N+1}\,,
\label{expod} \\
r &\simeq& \frac{16}{2N+1}\,,
\label{expo}
\ea
which correspond to taking the limit $n \to \infty$
in Eqs.~(\ref{chao2d}) and (\ref{chao2}).

In Fig.~\ref{fig4} we plot the theoretical values of 
$n_{\rm s}$ and $r$ for $N=50, 60, 70$ as well as
the 1$\sigma$ and 2$\sigma$ observational contours.
This shows that even the steep exponential potentials 
with $\beta^2 \gg 1$ are compatible with the current 
observational data.

Using the WMAP normalization ${\cal P}_{\rm s} \simeq 
2.4 \times 10^{-9}$ at $N=60$ in the regime 
$V/(M^2 M_{\rm pl}^2) \gg 1$, we obtain 
the constraint
\be
\beta \simeq
8.8 \times 10^{-6}\,\frac{M_{\rm pl}}{M}\,.
\label{lamcon}
\ee
For $M/M_{\rm pl} \ll 10^{-5}$ one has 
$\beta \gg 1$.

\subsection{Natural inflation}

Natural inflation \cite{Freese} is described by the potential 
\be
V(\phi)=\Lambda^4 \left[ 1+\cos \left( \frac{\phi}
{f} \right) \right]\,,
\label{natural}
\ee
where $\Lambda$ and $f$ are constants having 
the dimension of mass.
In the absence of the field derivative coupling to gravity 
the above potential can be compatible with observational data 
only for $f \gtrsim 3.5 M_{\rm pl}$ \cite{Savage}.
Then a global symmetry associated with 
the pseudo-Nambu-Goldstone-boson is broken 
above the quantum gravity scale, in which case
standard quantum field theory may not be applicable.
If the potential (\ref{natural}) originates from the string 
axion, the regime $f \gtrsim M_{\rm pl}$ is not generally 
realized \cite{Banks}.
This problem can be circumvented by taking into 
account the field derivative coupling 
to gravity\footnote{See Refs.~\cite{axionlarge} for other attempts to realize
$f \lesssim M_{\rm pl}$ in large-field axion models.} \cite{Germani2}.

\begin{figure}
\includegraphics[height=3.3in,width=3.4in]{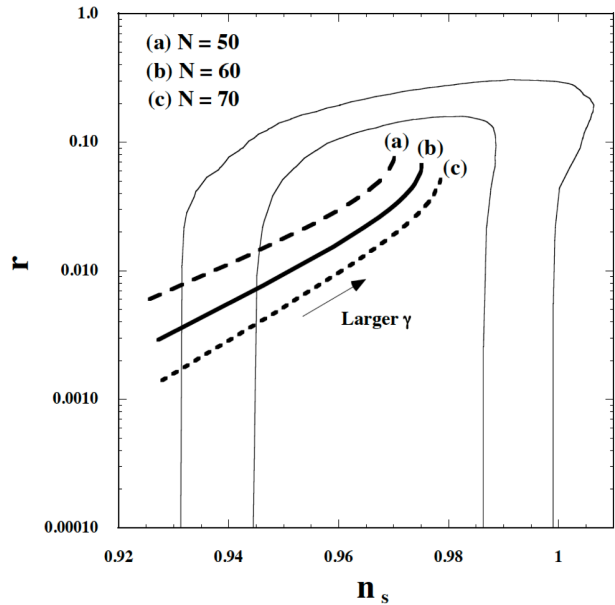}
\caption{Observational constraints on natural inflation
with the potential $V(\phi)=\Lambda^4 
\left[ 1+\cos \left( \phi/f \right) \right]$
in the range $7 \le \gamma=f^2 \Lambda^4/(M^2 M_{\rm pl}^4) 
\le 10^6$ with ${\cal A} \gg 1$
for three different values of $N$.
The $1\sigma$ and $2\sigma$ observational contours 
are the same as those in Fig.~\ref{fig2}.
For larger $\gamma$, both $n_{\rm s}$ and $r$
increase toward the values $n_{\rm s}=1-6/(4N+1)$
and $r=16/(4N+1)$.
\label{fig5}}
\end{figure}

In the following we focus on the case in which the condition 
${\cal A} \gg 1$ is satisfied during the whole stage of inflation 
($0<\phi< \pi f$).
{}From Eq.~(\ref{epV2}) the end of inflation is characterized by 
\be
\cos \chi_f
\simeq -1+\frac{\sqrt{16\gamma+1}-1}{4\gamma}\,,
\label{chif}
\ee
where $\chi_f=\phi_f/f$ and
\be
\gamma=\frac{f^2 \Lambda^4}{M^2 M_{\rm pl}^4}\,.
\ee
The number of e-foldings is given by 
\be
N \simeq \gamma \left[ {\cal F} (\chi_f)-{\cal F} (\chi) \right]\,,
\label{efoldnatural}
\ee
where $\chi=\phi/f$ and 
\be
{\cal F} (\chi)=2 \ln \left( \frac{1}{\sin \chi}
-\frac{1}{\tan \chi} \right)
+2\ln\,(\sin \chi)+\cos \chi\,.
\ee
The scalar spectral index and the tensor-to-scalar ratio are
\ba
n_{\rm s} &=& 1-\frac{2}{\gamma}\frac{2-\cos \chi}
{(1+\cos \chi)^2}\,,
\label{nsnatural} \\
r &=& \frac{8}{\gamma} \frac{1-\cos \chi}
{(1+\cos \chi)^2}\,.
\label{rnatural}
\ea
For given $\gamma$ the field value $\chi_f$ is known from 
Eq.~(\ref{chif}). 
Since $\chi$ is determined by Eq.~(\ref{efoldnatural})
for given values of $\gamma$ and $N$, we can numerically 
evaluate $n_{\rm s}$ and $r$ as functions of $\gamma$
for several different numbers of e-foldings.
In Fig.~\ref{fig5} we plot $n_{\rm s}$ and $r$ in the range 
$7 \le \gamma \le 10^6$, together with the 1$\sigma$
and $2\sigma$ observational contours.
For increasing $\gamma$, both $n_{\rm s}$ and 
$r$ get larger.

In the limit that $\gamma \gg 1$ it is possible to estimate
$n_{\rm s}$ and $r$ analytically.
{}From Eq.~(\ref{chif}) one has $\cos \chi_f
 \simeq -1+1/\sqrt{\gamma}$
and hence $\chi_f$ is close to $\pi$, as 
$(\pi -\chi_f)^4 \simeq 4/\gamma$.
Using Eq.~(\ref{efoldnatural}), it follows that 
$N \simeq (\gamma/16) (\pi -\chi)^4-1/4$.
Then Eqs.~(\ref{nsnatural}) and (\ref{rnatural}) 
reduce to 
\ba
n_{\rm s} &\simeq& 1-\frac{6}{4N+1}\,,
\label{nans} \\
r &\simeq& \frac{16}{4N+1}\,.
\label{nar} 
\ea
These values correspond to $n=2$ in 
Eqs.~(\ref{chao2d}) and (\ref{chao2}).
This comes from the fact that in the regime
$\gamma \gg 1$ inflation occurs around the 
potential minimum at $\phi=\pi f$.
As in the case of the quadratic potential 
$V(\phi)=m^2 \phi^2/2$ with $\alpha=m^2/M^2 \gg 1$, 
natural inflation with $\gamma \gg 1$ is compatible with 
the current observations.

For $N=60$ the parameter $\gamma$ is constrained to be 
\be
\gamma>7.5~~(95\,\%~{\rm CL})\,,
\ee
and $\gamma>9.8$ (68 \% CL).
For $N=50$ the constraints are
$\gamma>7.8$ (95 \% CL) and 
$\gamma>10.6$ (68 \% CL).
In the regime $\gamma \gg 1$ the WMAP normalization 
at $N=60$ gives 
\be
f \simeq 8 \times 10^4\, \gamma^{1/4}\,M\,.
\ee
If $M \lesssim 10^{-5} \gamma^{-1/4}M_{\rm pl}$, 
then one has $f \lesssim M_{\rm pl}$.

\subsection{Hybrid inflation}

Finally we study hybrid inflation with the potential 
\be
V(\phi)=V_0+\frac12 m^2 \phi^2\,,
\ee
where $V_0$ and $m$ are constants. 
Inflation ends at a bifurcation point given by
$\phi=\phi_c$ due to the appearance of 
a tachyonic instability driven by another field $\chi$.
As in the case of the original hybrid inflation \cite{hybrid} 
we focus on the regime $V_0 \gg m^2 \phi^2/2$.
Note that in another regime $V_0 \ll m^2 \phi^2/2$
the situation is similar to that in chaotic inflation
discussed in Sec.~\ref{chaoticsec}.

Using Eq.~(\ref{efold}) the field value can be estimated as
\be
\phi \simeq \phi_c \exp \left( \frac{\nu}{1+\mu} N \right)\,,
\ee
where $\mu$ and $\nu$ are positive constants defined by 
\be
\mu=\frac{V_0}{M^2 M_{\rm pl}^2}\,,\qquad
\nu=\frac{m^2 M_{\rm pl}^2}{V_0}\,.
\ee
{}From Eqs.~(\ref{ns}) and (\ref{ratio}) we obtain
\ba
n_{\rm s} &\simeq& 1+\frac{2 \nu}{1+\mu}\,,
\label{bybridns} \\
r &\simeq & \frac{8 \nu^2}{1+\mu} \left( \frac{\phi_c}{M_{\rm pl}}
\right)^2\,e^{(n_{\rm s}-1)N}\,,
\label{bybridr}
\ea
which mean that the scalar power spectrum 
is blued-tilted ($n_{\rm s} >1$).
Compared to the standard hybrid inflation, the presence 
of the field derivative coupling to gravity ($\mu>0$) 
leads to $n_{\rm s}$ close to 1 as well as the suppressed 
tensor-to-scalar ratio.
In the limit that $\mu \gg 1$ one has $n_{\rm s} \to 1$
and $r \to 0$, i.e. the Harrison-Zel'dovich (HZ) spectrum.
The HZ spectrum is under the observational pressure \cite{WMAP7}, 
but this property is subject to change depending on  
the assumptions about the reionization scenario \cite{reion}. 
The future high-precision observations will provide
a more concrete answer about this issue.

In the regime $\mu \gg 1$ and $\nu \ll 1$ one has $\phi \simeq \phi_c$ 
for $N \sim 60$.
{}From the WMAP normalization it follows that 
\be
\frac{\mu}{\nu} \frac{M}{\phi_c}
 \simeq 5.3 \times 10^{-4}\,.
\ee
If $\mu=10^2$, $\nu=10^{-2}$, $\phi_c=0.2 M_{\rm pl}$, 
for example, $M \simeq 10^{-8}$~GeV. In this case 
Eqs.~(\ref{bybridns}) and (\ref{bybridr}) give
$n_{\rm s} \simeq 1.0002$ and $r \simeq 3 \times 10^{-7}$, 
which is close to the HZ spectrum.

\section{Conclusions}

In this paper we have studied observational constraints on 
a number of representative inflationary models with a field 
derivative coupling to the Einstein tensor, i.e.
$G^{\mu \nu} \partial_{\mu} \phi \partial_{\nu} \phi/(2M^2)$.
Such a non-minimal derivative coupling has an asymptotic 
local shift symmetry for a slow-rolling scalar field
satisfying the condition $\varepsilon=(\partial \phi)^2
/(M^2 M_{\rm pl}^2) \ll 1$.
Since the strong coupling scale of the theory is 
around the Planck scale for $\varepsilon \ll 1$, 
quantum corrections to the inflaton potential 
can be suppressed during inflation.

The non-minimal derivative coupling to gravity leads to 
a gravitationally enhanced friction for the scalar field.
This property allows us to accommodate steep potentials 
with theoretically natural model parameters for realizing inflation.
Not only the quartic potential 
$V(\phi)=\lambda \phi^4/4$ with $\lambda \sim 0.1$
but the potential $V(\phi)=\Lambda^4 [1+\cos (\phi/f)]$
with $f  \ll M_{\rm pl}$ gives rise to 
cosmic acceleration consistent with the amplitude of 
the CMB temperature anisotropies.
Moreover the exponential potential $V(\phi)=
V_0 e^{\beta \phi/M_{\rm pl}}$, which often appears
after the compactification of extra dimensions in higher-dimensional 
theories, can lead to inflation even for $\beta^2 \gg 1$
in the presence of the field derivative coupling to gravity.

For the potential $V(\phi)=(\lambda/n)\phi^n$ of chaotic inflation
the tensor-to-scalar ratio $r$ decreases for larger 
$\alpha=\lambda M_{\rm pl}^{n-2}/M^2$, whereas the 
scalar spectral index $n_{\rm s}$ increases.
In the limit that $\alpha \to \infty$, $n_{\rm s}$ and $r$
approach the values given in Eqs.~(\ref{chao2d}) and
(\ref{chao2}).
As we see in Figs.~\ref{fig2} and \ref{fig3}, for both $n=2$ and $n=4$, 
the asymptotic values of $n_{\rm s}$ and $r$ are within the $1\sigma$ 
observational bound derived
by the joint data analysis of WMAP7, BAO, and HST.
For the quartic potential with $N=60$ we found that the parameter
$\alpha=\lambda (M_{\rm pl}/M)^2$ is constrained 
to be $\alpha>3 \times 10^{-5}$
at the 95 \% confidence level.

For the exponential potential $V(\phi)=V_0 e^{\beta \phi/M_{\rm pl}}$
the asymptotic values of $n_{\rm s}$ and $r$
in the regime $V/(M^2 M_{\rm pl}^2) \gg 1$ are given by 
Eqs.~(\ref{expod}) and (\ref{expo}), with $\beta$ 
constrained as Eq.~(\ref{lamcon}) from the WMAP normalization.
Figure \ref{fig4} shows that steep exponential potentials 
with $\beta^2 \gg 1$ can be compatible with the current 
observational data.

In natural inflation with the potential $V(\phi)=\Lambda^4 [1+\cos (\phi/f)]$
the observables can be parametrized by the parameter 
$\gamma=f^2 \Lambda^4/(M^2 M_{\rm pl}^4)$ in the regime 
$V/(M^2 M_{\rm pl}^2) \gg 1$.
For larger $\gamma$, both $n_{\rm s}$ and $r$ tend to increase
toward the values given in 
Eqs.~(\ref{nans}) and (\ref{nar}).
These asymptotic values correspond to those derived in 
Eqs.~(\ref{chao2d}) and (\ref{chao2}) for $n=2$ with $\alpha \gg 1$.
This property comes from the fact that for $\gamma \gg 1$
inflation occurs around the potential minimum at $\phi=\pi f$.
The observational bound on the parameter $\gamma$ is
found to be $\gamma>7.5$ for $N=60$ (95 \% CL).
{}From the WMAP normalization the symmetry breaking scale 
$f$ is constrained to be $f \simeq 8 \times 10^4\,\gamma^{1/4} M$, 
which can be smaller than $M_{\rm pl}$ for 
$M \lesssim 10^{-5} \gamma^{-1/4}M_{\rm pl}$. 

In hybrid inflation with the potential $V(\phi)=V_0+m^2 \phi^2/2$
(where $V_0 \gg m^2 \phi^2/2$) the field derivative coupling to gravity 
leads to the blue-tilted scalar power spectrum close to $n_{\rm s}=1$.
Compared to standard hybrid inflation, the power spectrum approaches
the HZ one, i.e. $n_{\rm s}=1$ and $r=0$.
The HZ spectrum is in tension with observations, but we have to
caution that this property is affected by the assumption of 
the reionization scenario.

If future observations such as PLANCK \cite{PLANCK} can 
constrain the tensor-to-scalar ratio at the level of $r \lesssim {\cal O}(0.01)$, 
this will allow us to place tighter bounds on the 
inflationary models with the field derivative coupling to gravity.
We hope that we can discriminate between a host of 
inflationary models within the next few years.

\section*{ACKNOWLEDGEMENTS}
This work is supported by the Grant-in-Aid for
Scientific Research Fund of the Fund of the 
JSPS No 30318802 and Scientific Research 
on Innovative Areas (No.~21111006). 
The author thanks Antonio De Felice for warm hospitality 
during his stay in Naresuan University.
The author is also grateful to Cristiano Germani
for useful correspondence. 

\end{document}